\documentclass[a4,10pt]{article}
\usepackage{amsmath,amsfonts,amssymb,graphicx,epsfig}
\usepackage{colordvi}
\def\ba{\begin{eqnarray}}
\def\ea{\end{eqnarray}}

\begin{document}
\title{\textbf{How hole defects modify vortex dynamics in ferromagnetic nanodisks}}
\author{W.A. Moura-Melo\thanks{E-mail address: winder@ufv.br}, A.R. Pereira\thanks{E-mail address: apereira@ufv.br}, R.L. Silva, and N.M. Oliveira-Neto\\ \small\it Departamento de F\'{\i}sica, Universidade Federal
de Vi\c{c}osa,\\ \small\it 36570-000, Vi\c{c}osa, Minas
Gerais, Brazil \\ }

 \date{\today}
\maketitle
\begin{abstract}
Defects introduced in ferromagnetic nanodisks may deeply affect the structure and dynamics of stable vortex-like magnetization. Here, analytical techniques are used for studying, among other dynamical aspects, how a small cylindrical cavity modify the oscillatory modes of the vortex. For instance, we have realized that if the vortex is nucleated out from the hole its gyrotropic frequencies are shifted below. Modifications become even more pronounced when the vortex core is partially or completely captured by the hole. In these cases, the gyrovector can be partially or completely suppressed, so that the associated frequencies increase considerably, say, from some times to several powers. Possible relevance of our results for understanding other aspects of vortex dynamics in the presence of cavities and/or structural defects are also discussed.  
\end{abstract}

{\bf Pacs:} 75.75.+a; 75.70.-i; 75.70.Rf  \\


\section{Introduction and Motivation}

\indent Ferromagnetic Permalloy nanodisks with lateral dimension (thickness, $L$) around some dozens of nanometers have been
fabricated and investigated for their potential applications in a
number of magnetoelectronic mechanisms. In particular, it has
been observed that above the so-called single-domain limit,
magnetic vortex states appear in these samples\cite{Cowburn,Raabe,Ross}, exhibiting a
planar-like arrangement of spins outside the core, where a
perpendicular magnetization is observed\cite{Shinjo,Miltat,Wacho}.
As long as one could manipulate these states other possibilities
would emerge. In fact, one way towards this control is obtained by
removing some small portions of the magnetic nanodisk, in such a
way that the defects (cavities) so created work by attracting and
eventually pinning the vortex around themselves \cite{RahmPRL95,RahmAPL82,RahmJAP95,AfranioPRB2005,AfranioJAP2005,Capvortex,AfranioJF,FagnerPLA2004} (similar effect also applies to soliton-like spin excitations \cite{nossoPRB2003}).
Based upon such an idea, Rahm and coworkers \cite{RahmAPL85} have studied the cases of two, three and four holes (each
of them with diameter $\sim85\,{\rm nm}$) inserted in a disk with
diameter $\sim 500\,{\rm nm}$, separated by around $150\,{\rm
nm}\,-\,200\,{\rm nm}$. Their experimental results confirmed the previous statement about vortex pinning and put forward the possibility of using these stable states as serious candidates for magnetic memory and
logical applications as long as we could control vortex position,
for example, applying a suitable external magnetic field which
should shift the vortex center from one defect to another, and
vice-versa. Basic logical operations have been obtained in these lines by means of bistable magnetic switching\cite{RahmAPL87}.\\

When the vortex experiences a suitable alternating field its gyrotropic mode may be resonantly excited in the subGHz range, which has attracted a great deal of efforts for it is the lowest translational vortex mode\cite{G-ParkCrowellPRB67,GuslienkoAPL2002,G-GusJAP91-02,Usov,G-ZaspelIvanovJAP95-04,G-GusPRB72-05,G-GusPRL96-06}. Such a frequency depends upon the geometry of the nanodot and, in the case of thin cylindrical samples, it was initially believed that the aspect ratio, $L/R$, completely determines its values \cite{GuslienkoAPL2002,G-GusJAP91-02}. However, Park and Crowell \cite{ComptonCrowellPRL2006} have observed that crystalline defects working as very small pinning sites play a crucial role in the vortex dynamics, namely in the critical field that resonates the gyrotropic mode. Hoffman {\em et al} \cite{HoffmanPRB2007} have also observed a remarkable asymmetry in the magnetization reversal mechanism which they credited to such small defects. Kuepper {\em et al}\cite{Kuepper} have studied this mode when a (cylindrical) cavity is intentionally introduced in the sample. Their results clearly show how these cavities affect vortex motion as a whole, particularly its gyrotropic frequencies. In addition, Hoffmann {\em et al} \cite{HoffmanPRB2007} have verified that the two magnetostatic spin-wave modes degenerate to a single frequency when the vortex core is captured by a hole. Here, we shall show that a number of facts, including some ones observed by Kuepper {\em et al} \cite{Kuepper}, may be well understood in terms of very basic physical properties of the vortex structure and dynamics in the presence of a cavity. Qualitatively, our arguments and results may be useful for other similar systems were pinning sites/defects concern.

\section{The analytical model and basic results}

Consider a magnetic dot represented by a
small cylinder of radius $R$ and thickness $L$ (so that its aspect
ratio $L/R <<1$). Assume that along the
axial direction ($z$-axis), the magnetization $\vec{M}$ is
uniform; assume also negligible anisotropy energy (e.g., Permalloy samples). Furthermore, if we introduce $N$ isolated holes (each of
them with height $L$ and radius $\rho_N<<R$) in the dot, the total
magnetic energy of the nanodisk can be approximated, in the
continuum limit, by\cite{nossoJAP2007} 
\ba E_{\rm total}=\frac{L}{2}\int\int_D
\left[A(\partial_\mu\vec{m})\cdot(\partial^\mu\vec{m}) -M^2_s\,
\vec{m}\cdot(\vec{h}_m +2\vec{h}_{\rm ext})\right]
\prod^N_{i=1}\,U_i(\vec{r}-\vec{d}_i)\, d^2r \,, 
\ea
where $A$ is
the exchange coupling, $D$ is the area of the cylinder face,
$\vec{m}=\vec{M}/M_s$ is an unity vector describing magnetization
along $D$ ($M_s$ is the saturation magnetization),
$\vec{h}_m=\vec{h}_m(\vec{m})\equiv \vec{H}_m/M_s$ is the
demagnetizing field, $\vec{h}_{\rm ext}$ is an applied magnetic
field (Zeeman term) and $\mu=1,2$. The potential $U$, in turn,
brings about the effect of the holes distributed throughout the
nanodisk, say,
$\prod^N_{i=1}\,U_i(\vec{r}-\vec{d}_i)=U_1(\vec{r}-\vec{d}_1)\,U_2(\vec{r}-\vec{d}_2)\ldots
U_N(\vec{r}-\vec{d}_N)$, with 
\ba
 U_i(\vec{r}-\vec{d}_i)=\left\{\begin{array}l 0 \quad{\mbox{ \rm if}}\quad
\mid \vec{r}-\vec{d}_i \mid <\rho\\
                              1 \,\quad{\mbox{ \rm if}} \quad \mid \vec{r}-\vec{d}_i
                                 \mid \geq \rho  \end{array}\right.\,. \label{Uimp}
\ea
Therefore, the system of a dot with $N$ isolated antidots may
be viewed as a cylinder of radius $R$ and thickness $L$ with $N$
smaller cylindrical cavities with radius $\rho<<R$, each of them
centralized at $\vec{d}_i$. Here, we shall study the case $N=1$ (the treatment for $N\ge2$ may be performed in the same
way).\\
   
Let us sketch the framework: consider a cylindrically symmetric vortex-like
magnetization throughout the dot face, say, with the vortex core
centralized at $\vec{r}=\vec{0}$ which is the magnetization ground state of a
nanodisk, with $L/R<<1$, in the absence of both hole and external field. It is convenient to write $\vec{m}=(\sin\theta\cos\varphi,
\,\sin\theta\sin \varphi, \, \cos\theta)$, with
$\theta=\theta_v(r)$ and $\varphi=\arctan(y/x)\pm\pi/2$. The
function $\theta_v(r)$ may be approximated by $\sin\theta_v(r)=0$
in the dot center ($\vec{r}=\vec{0}$), while $\sin\theta_v(r)\to1$
far away the center, $\mid\vec{r}\mid=r>>l_{\rm ex}$ ($l_{\rm ex}=\sqrt{{A}/2\pi M_{s}}$ is the exchange
length). In words, the magnetization consists of a small
core where spins develop out-of-plane components for regularizing
the exchange energy, and an outer region where spins are
practically confined to the dot plane face, so minimizing the (stray) magnetostatic cost. In this ground-state, the
magnetic superficial charges in the lateral face of the dot and
the magnetic volumetric charges, $\vec{\nabla}\cdot\vec{m}$,
identically vanish yielding no contribution to $\vec{h}_m$. The
cavity affects this picture as follows: it yields a less exchange energy for the vortex, attracting it and eventually deforming its profile, according to the potential below\cite{AfranioPRB2005,AfranioJAP2005,Capvortex}:
\ba \label{Vex}
V_{\rm ex}\cong \frac{\pi AL}{2} \ln\left[ \left(1-s^2 \right) \left(1-\frac{\rho^2}{(sR -d)^2 +b^2}\right)\right]\,,
\ea
where $\vec{s}=\vec{r}/R$ is the relative vortex center displacement, $d$ is the distance between hole and disk centers, and $b$ is a constant introduced to avoid spurious divergences whenever the vortex is centralized at the hole, i.e., $(sR-d)=0$ (to be estimated later). Note that the first term inside $\ln$-function accounts for the deformation of the vortex while it dislocates from the center while the remaining one takes into account the hole effect. For ensuring the strict validity of the above and other contributions to the total energy we must consider $s$ sufficiently small, so that $d$ and $\rho$ should be also small compared to the disk radius, $R$.\\

In addition, the
distribution of magnetic charges throughout the internal edges of
the hole and mainly along the external lateral face of
the nanodisk (if the cavity is not at the center of the disk) increases the magnetostatic energy due to a
change in the product $\vec{m}\cdot\hat{n}_{s}$ ($\hat{n}_{s}$ are
unit vectors normal to the external lateral surface of the disk and
internal surfaces of the cavity). Thus, the demagnetizing field,
$\vec{h}_{m}$, can be obtained from its associated potential
$\Phi_m=\Phi_{\rm V}+\Phi^e_{\rm edge}+\Phi^i_{\rm edge}$, in the
usual way, $\vec{h}_m=-\vec{\nabla}\Phi_m$. Here, $\Phi_{\rm V}$
is the magnetostatic potential related to the volumetric charges,
while $\Phi^e_{\rm edge}$ and $\Phi^i_{\rm edge}$ come about from
the surface charges on the external and internal (hole) edges,
respectively. The contributions of the volumetric potential can be
neglected since the approximations assumed above lead to
$\vec{\nabla}\cdot\vec{m}=0$. Thus, the magnetostatic contribution reads\cite{AfranioPRB2005, AfranioJAP2005,Capvortex,nossoJAP2007}:
\ba \label{Vmag}
V_{\rm mag}\cong 2\pi^2 M^2_s\, L(R^2 -\rho^2)\left[ F_1(L/R)s^2  +\alpha F_1(L/\rho) \,(s-d/R)^2 +\beta F_1(L/\rho)\,(\rho/R)^2\right]\,,
\ea
where $F_1(x)= \int^\infty_0 \, dt [ J^2_1(t)\,(e^{xt} +xt -1)]/xt^2$. In addition, $(\alpha,\beta)=(1,0)$ for $|s-d/R|< \rho/R$ (vortex center is inside the hole), or $(0,1)$ if $|s-d/R|\ge \rho/R$ (out from the hole).\\

\begin{figure}[h!t]
\centering 
\includegraphics[angle=-90,width=12cm]{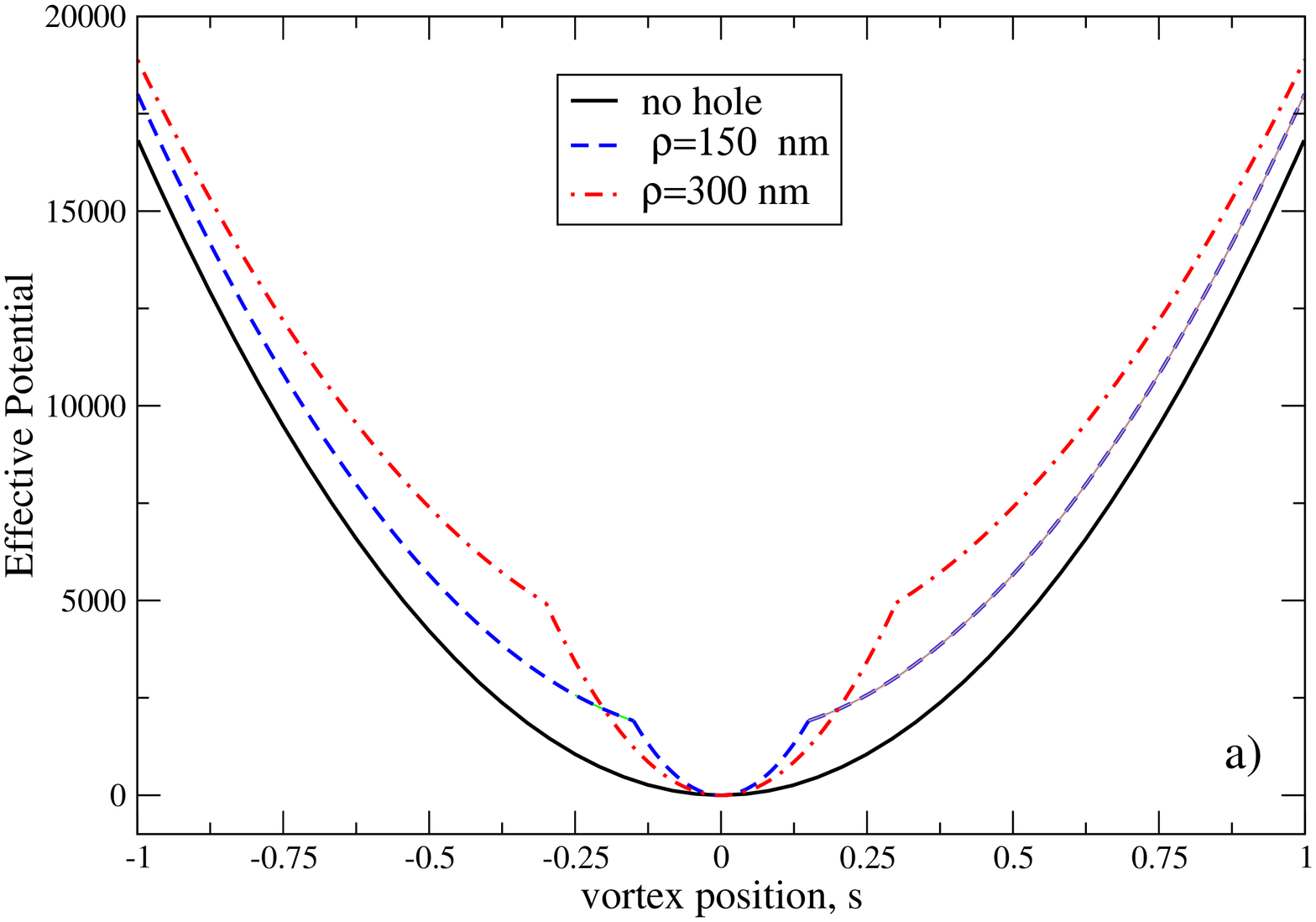}  \hskip .0cm
\vskip 1.5cm \hskip -.8cm
      \includegraphics[width=10cm]{figure1b.eps}  \caption{} \label{figure1}
\end{figure}

In the presence of an external field the following Zeeman term must be taken into account:
\ba \label{VZeeman}
V_{\rm Z}=-\pi\,L(R^2- \rho^2)h M^2_s \, s+ {\cal O}(s^3)\,,
\ea
with $h=|\vec{H}|/M_s$. Therefore, the effective potential experienced by the vortex reads $V_{\rm eff}(s)=  V_{\rm ex}+V_{\rm mag}+V_{\rm Z}$. Depending on the relative hole-vortex centers we have the following possibilities: i) the hole is centered at the disk, $\vec{d}=(0,0)$, yielding a unique point that minimizes energy, $\vec{s}_0=(0,0)$; ii) the hole center is at $\vec{d}=(X,),\, X>0$ (without loss of generality) and the vortex center is far apart $X$, say, $\alpha=0$, so that we again have $\vec{s}_0=(0,0)$ as the unique equilibrium position (the hole practically does not affect the vortex), and; iii) $d=X>0$ and the vortex center is inside the hole (or very close to it), $\alpha =1$, then besides $\vec{s}_0$ it also experiences a new EP at $\vec{s}_1=(x_1/{R},0)$, with
\ba 
\frac{x_1}{R}=\frac{4(R^2-\rho^2)F_1(L/\rho) - l^2_{\rm ex} \ln \left(1-\frac{\rho^2}{b^2} \right)}{4(R^2-\rho^2)\left[F_1(L/R) +F_1(L/\rho) \right] - l^2_{\rm ex}\left[1+ \ln\left(1-\frac{\rho^2}{b^2}\right)\right]}\frac{X}{R} \,,
\ea
which depends on the disk and hole relative geometries and is located between the disk and hole centers. Indeed $x_1<X$ by virtue of the magnetostatic effect which always tends to centralize the vortex in the disk (these cases are illustrated in Figure \ref{figure1}). The parameter $b$ may be estimated by equating the vortex exchange energy in the presence of the hole, eq. (\ref{Vex}), to its normalized rest energy, $E_0=2\pi AL\ln(R/l_0)$, in the same situation, $E_{0\, {\rm hole}}=2\pi AL\ln(R/\rho)$, say, $V_{\rm ex}=E_{0\, {\rm hole}} -E_0 -E_{\rm core}$, where $E_{\rm core}=2\pi AL$ is the core energy, which must be taken into account if the vortex is out from the hole. After some algebra and assuming $l_0\cong l_{\rm ex}$ and $\rho\ge l_{\rm ex}$, we get:
\ba \label{b}
\ln\left(1-\frac{\rho^2}{b^2}\right)=-4\left[ 1+\ln(\rho/l_{\rm ex})\right]\,,
\ea
which gives $b\gtrsim \rho$ (in the thermodynamical limit $b\cong 1.04 \rho$ \cite{AfranioJF}).\\

\begin{figure}[h!t]
\centering \hskip 1cm 
{\includegraphics[angle=-90,width=10cm]{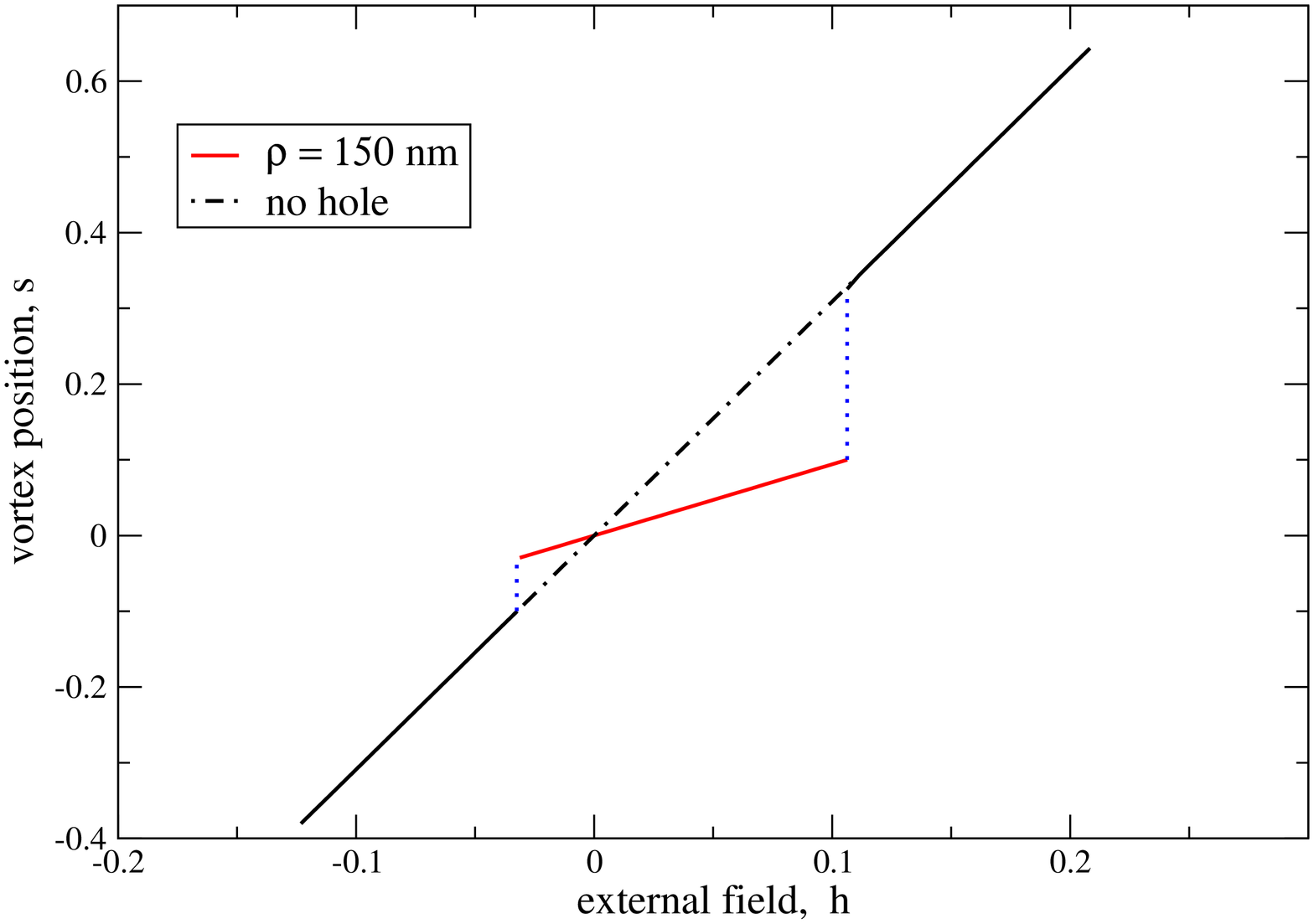}}
\caption{} \label{figure2}
\end{figure}

Therefore, each non-centered defect may provide an additional EP for the vortex center, as observed in experiments\cite{RahmPRL95,RahmAPL82,RahmJAP95} and predicted theoretically \cite{AfranioPRB2005, AfranioJAP2005,Capvortex,nossoJAP2007}. Namely, note that if $\rho>\rho_{\rm cr}$ then the potential well at the hole is deeper than that at the disk center (see Fig. \ref{figure1}.b). In this situation, once the vortex center is captured by the cavity it will remain there unless a strong enough perturbation (like an external homogeneous field) take it to another EP. Indeed, turning $h$ on the vortex center is shifted from its old EP, along $x$, by:
\ba 
\frac{x_h}{R}= \frac{\pm h(R^2-\rho^2)}{4\pi(R^2-\rho^2)\,\left[F_1(L/R)+\alpha\,F_1(L/\rho) \right] -2\pi  l^2_{\rm ex} \left[1 +\alpha \ln\left(1+\frac{\rho^2}{b^2} \right)\right]}\,,
\ea
for $\vec{h}=\pm h\hat{y}$ and clockwise vortex orientation. Thus, the vortex EP's with both hole and external field are $\vec{s}_{0,h}=(x_h,0)$ and $\vec{s}_{\rm 1,h}=(x_1+x_h, 0)$. Furthermore, once the hole attracts the vortex to its center, there is an asymmetry in the
vortex path as $h$ is varied, say, from
-1 to +1. At fields $\sim -1$ the vortex is practically
annihilated at the disk border. As the field is gradually removed
towards zero, the vortex continuously displaces to the disk
center. However, when it gets close to the hole border it abruptly
jumps inside it, at a critical field, $h_{\rm cr1}<0$, nucleating
around the hole center. For a wide range of the field the vortex
center remains inside the hole. Only at a sufficiently strong
field, $h_{\rm cr2}>0$, the vortex center is released from the
hole, performing another abrupt jump (note that $|h_{\rm cr2}|>|h_{\rm cr1}|$). This scenario is depicted in
Figure \ref{figure2}, where the usual linear path is also presented for comparison. These should be compared to those obtained in the experiments of Ref.\cite{RahmPRL95}, its Figures 3 and 4 and related text; namely note that our results fit qualitatively well those reported in this work.\\

In addition, there follows that the spring-like constant experienced by the vortex center reads:
\ba
 k= \pi M^2_s L \left(\frac{R^2-\rho^2}{R^2}\right) \left[ 4\pi\left[F_1(L/R)+\alpha\,F_1(L/\rho)\right] -\frac{l^2_{\rm ex}}{(R^2 -\rho^2)} \left[1-\alpha \,+\alpha \ln\left(1-\frac{\rho^2}{b^2}\right) \right)\right] \label{khole}\,,
\ea
which is clearly enhanced whenever the vortex center is inside the hole, $\alpha=1$. Results above are strictly valid within the rigid vortex assumption, and might be extended, for instance, to the two side charges regime.\\

Besides the spring-like constant, the hole may also lead to profound modification in the vortex structure and dynamics. For instance, like an external field a hole may trigger the formation of non-homogeneous magnetization pattern at the disk and hole borders \cite{AfranioPRB2005,AfranioJAP2005,Capvortex}; also, when its center is captured by the hole its gyrotropic motion is suppressed\cite{Kuepper}. Other possibilities will be discussed in what follows. It should be stressed that in the absence of the hole our results recover those already found in the literature \cite{GuslienkoAPL2002}.\\

\section{Further results and discussion}
Consider two disks, B and C, with radii $R_B$ and $R_C$. Disk B has a non-centered hole, of radius $\rho_B$ at $\vec{d}_B=(d,0),\, d>0$, while in disk C the hole, $\rho_C$, is centered. Let A denote a disk, $R_A$, without hole. Their thickness are $L_A$, $L_B$, and $L_C$. For concreteness we also assume typical Permalloy  parameters: $A=1.3\,{\rm x}\,10^{-6}\,{\rm erg/cm}=1.3\,{\rm x}\, 10^{-11} \,{\rm J/m}$ and $M_s=800\,{\rm emu/cm^3}=800 \,{\rm KA/m}$, so that the exchange length, $l_{\rm ex}=\sqrt{A/2\pi M^2_s}\simeq 5\,{\rm nm}$. When vortex gyrotropic vector concerns, its gyroratio is $\gamma/2\pi =LM_s/2G= 2.95 \,{\rm GHz/KOe}=37\,{\rm KHz\,m/A}$.\\

\begin{figure}[h!t]
\centering 
{\includegraphics[angle=-90,width=12cm]{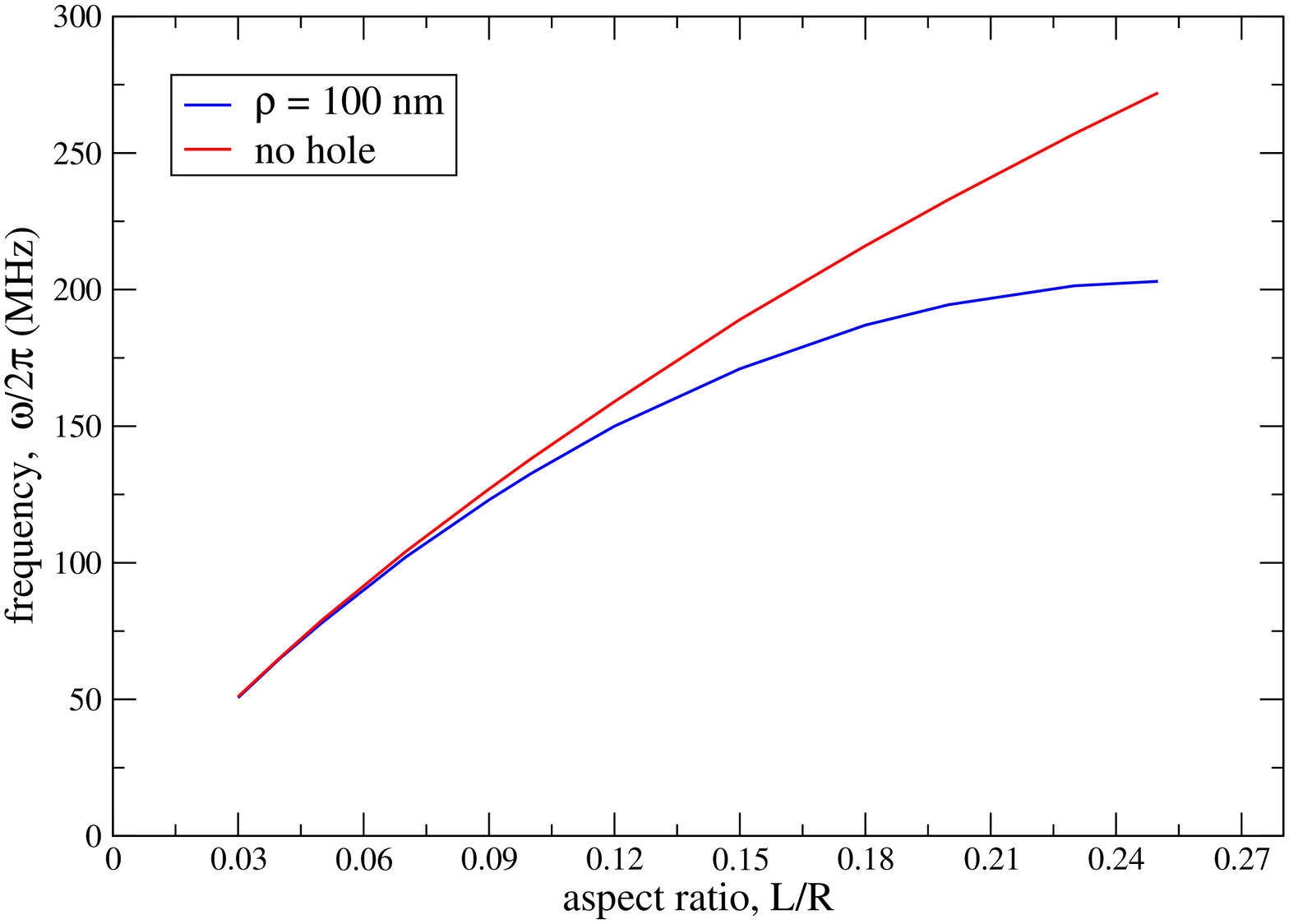}} \caption{} \label{figure3}
\end{figure}

Disks B and C match their counterparts 2 and 3 from Ref.\cite{Kuepper}, with $R_2=1\,{\rm \mu m}$, $\rho_2=300\,{\rm nm}$, and $d$ is assumed $\cong 400\,{\rm nm}$ (not provided in Ref.\cite{Kuepper}; we have estimated it by means of the available images); $R_3=1.5 \,{\rm \mu m}$ and $\rho_3=300\,{\rm nm}$, respectively. Our disk A matches their disk 1, with $R_1=750\,{\rm nm}$; all the three disks are $L=50\,{\rm nm}$ thick. In their work, the authors have estimated (experimental values are not presented) the gyrotropic frequency of the vortex in their samples, based upon the two-vortex side charges (without hole effects), to be $333.5\,{\rm MHz}$, $249\,{\rm MHz}$, and $164.5\,{\rm MHz}$, for disks 1, 2 and 3, respectively. Once the values above are believed to superestimate experimental findings in $10\%-15\%$, hole effects are expected to account for such deviations, say, in hollowed samples, like disks 2 and 3. As usual, such frequencies are given by $\omega_G=k/G$, with $G=2\pi M_sLqp/\gamma=$, where $q$ and $p$ account for the chirality and polarization of the vortex (the possibility of fractional polarization is treated below). Once the vortex is nucleated out from the hole, its gyrovector is completely turned on but $k$ changes according eq. (\ref{khole}), with $\alpha=0$. In this case the exchange contribution to $k$, third term in eq. (\ref{khole}), is very small compared to the magnetostatic one for these disks, so that the main hole contribution to the gyrofrequency comes from the factor $(R^2-\rho^2)/R^2$ which reads $0.91$ and $0.96$ for disks 2 and 3, respectively. Thus, their respective frequencies are decreased by about 9\% and 4\%, in the presence of the hole. Such deviations appear to become larger as $L/R$ increases as shown in Fig. \ref{figure3}.\\

These are slight effects if compared to those in which the vortex center is captured by the hole, suppressing its gyrovector, once its core is removed. In this case, the vortex dynamics is deeply affected and low frequency modes (like the usual gyrotropic) no longer take place, as observed in experiments \cite{Kuepper}. However, other oscillations are possible within the dynamical equation:
\ba \label{LandauLifshitz}
{\bf G}\times \dot{\bf x}- k{\bf x}=M_v \ddot{\bf x}\,,
\ea
where ${\bf x}$ and $M_v\cong \pi\hbar^2 \ln(R^2/l^2_0)/8AL l^2_{\rm ex}$ are the vortex position and mass ($\dot{\bf x}\equiv d{\bf x}/dt$, etc). This mass expression has been adapted from Ref. \cite{Wysin1}, with $l_0$ being the core radius. Instead of $\ln(R^2/l^2_0)$ a $R^2/l^2_0$ behavior has been recently claimed to better fitting  simulations, at least for planar vortices\cite{AfranioJF}. Indeed, additionally to the lower gyrotropic, $\omega_G=k/G$, there is an harmonic vortex oscillation around the hole with frequency $\omega_M=\sqrt{k/M_v}$, namely if the vortex is inside the hole. For typical samples, like those considered above, we find $M_v\cong 0.01- 1\, m_e$ (using $\ln(R^2/l^2_0)$ and $R^2/l^2_0$, respectively; $m_e$ is the electronic mass), so that harmonic oscillations take place around $\omega_H\cong 10^{15} - 10^{16}\, {\rm Hz}$ ($\sim 10^6$ times the gyrotropic ones), which is much higher than currently available experimental capabilities, around some dozens of GHz \cite{KuepperPC}.\\

Small vortex oscillations around a given EP naturally take place for balancing net energy of the magnetization, for instance, against an external alternating field. When this occurs some portion of the vortex central region can escape from the hole and possibly develops out-of-plane spin components, so developing part of the gyrovector. Such a gyrovector (lying, probably, at or above the GHz scale, like below) could be observed if a suitable alternating field were applied to resonantly turn on such a mode. In this case, additionally to the harmonic oscillation perpendicular to the field the vortex would also orbit around the defect (like it usually does around an EP). This would ensure that, whenever taken away from the hole, vortex gyrovector is developed from the core border towards the center, demanding gradual and continuous flipping of the normal magnetization. [Conversely, if no such additional orbiting motion were observed, then we should expect that vortex core (and gyrovector) starts to be formed from the center to the border, a fact less probable energetically, once it demands an abrupt $\pi/2$-flipping of the spin at the center]. Therefore, such experiments could also reveal the fractional gyrovector structure of this vortex-like configuration, provided by the dynamics of a vortex captured by a hole. Indeed, it should be stressed that those spin sites at the hole border (like those at the disk border) have a link topology different from those located at the bulk: they have only five nearest-neighbor against six from the bulk sites. Such a asymmetry could somewhat induce, at a tick strip around the hole border, out-of-plane spin component, for instance, through exchange anisotropy. In this case, a fractional gyrovector takes place, like below:
\ba \label{fracG}
\vec{G}_{\rm frac}=-\frac{M_s L}{\gamma} \int^{R}_{0} (\nabla\cos\Theta_\chi)\times (\nabla\Phi)\, d^2 r= \chi\vec{G} \,,
\ea 
where the integral is effectively evaluated along a small distance, $\delta$, from the hole border, $r'=r-\rho=0$, towards the bulk, where $\nabla\Theta_\chi\neq0$. We have taken $\Phi=\pm\phi \pm \pi/2$, $\phi$ is the azimuthal angle on the disk plane, and $\Theta_\chi$ is such that $\cos\Theta_\chi\to \chi$ as $r\to\rho$, $|\chi|<1$, and $\cos\Theta_\chi\to0$ for $r>\rho+\delta$. A trial solution is given by $\cos(\Theta_\chi)=\mp\chi \{[(r-\rho)^2 -\delta^2]/[(r-\rho)^2 +\delta^2]\}^n$, with $n$ a positive real parameter (namely, $n=4$ yields a smoother behavior for the magnetization \cite{BahianaNanotech2007}). In words, $\Theta_\chi$ describes a magnetization configuration which presents a fractional polarization at the hole border, $p_\chi=\chi$, falling off rapidly as we go towards disk border. Thus, $\chi \in(0,1)$ accounts for how much the spins around the hole border are flipped towards the normal. Indeed, such a flipping is expected to be smaller as the hole increases, so that above a critical value no net magnetization perpendicular to the disk plane takes place anywhere. In the case of a static vortex this critical hole size goes around $\sim 0.3 l_{\rm ex}$ \cite{ShekaJMMM2007}, but could be largely increased in the dynamical case, once the vortex center can now moves towards hole border and even escape outside. Therefore, whether this fractional polarization actual takes place seems to strongly depend on the relative location of the vortex center to the hole border at a given time, varying according the distance between them changes. However, if the hole is very large the displacement of the vortex center to the hole border may be highly energy costing leading eventually to a deep deformation in its profile or even its annhilation, so that its central region is kept inside the hole yielding no net polarization at all. This seems to be the case of sample 3 studied in Ref. \cite{Kuepper}, whose observations have led the authors to conclude that the gyromode is completed suppressed, once the vortex core is captured by the centralized hole. Nevertheless, just for comparison, consider that the gyrovector was not completely turned off, but was reduced to, say, $10\%$ ($\chi=0.1$, i.e., maximum spin flip, $\Theta_{\chi}|_{\rm max}=9^o$). In this case, sample with $\rho=300\,{\rm nm}$ and $R=1,5 \,\mu{\rm m}$, the frequency associated to the fractional gyrovector reads  
\ba\label{omegafrac}
\omega_{\rm frac}=k/G_{\rm frac}\sim 10 \,{\rm GHz}\,,
\ea
increasing as $\chi$ is diminished. Even in this situation of fractional gyrovector our analysis may explain the reason why in the work of Ref.\cite{Kuepper} no trace of the subGHz gyrotropic mode was observed: the applied field frequency is far below that required for resonantly excite the fractional mode, as eq. (\ref{omegafrac}) predicts.\\

\section{Conclusions and Prospects\\}
We have realized that the introduction of a hole into a ferromagnetic nanodisk, with small aspect ratio, may deeply modify the structure and dynamics of a vortex-like magnetization. Such effects come about once a hole change the effective potential experienced by the vortex in such a way that the vortex center is attracted towards the hole. As a vortex is captured by a hole its out-of-plane components, accounting for its polarization, are greatly diminished or even vanished. Such a capture may be clearly viewed in the plot of the vortex center position against a uniformly varying external field, Figure \ref{figure2}. A moving vortex whose center is not inside a hole experiences a decreasing in its associated gyrotropic frequency, which may become pronounced as the disk and hole sizes are comparable.\\

We have also raised the interesting possibility of the fractional polarization of moving vortex. This would happens if the vortex center were inside the hole, but sufficiently close to its border, so that some portion of the out-of-plane magnetization show up. Among other issues, its observation could answer how the vortex core is formed as it is released from a hole. Our results could have some relevance to the study of structural defects in nanomagnets, once a analytical model for understanding such structures in these frameworks is still lacking.\\

A very interesting result to be addressed is the analysis of the vortex motion in the presence of the hole and/or external field. Analytically, this task demands the resolution of complicated differential equations obtained from eq.(\ref{LandauLifshitz}) with the force, ${\bf F}=M_v \ddot{\bf x}$, given by ${\bf F}=-{\bf \nabla}V_{\rm eff}$. Simulations have revealed the richness of such dynamics, including the possibility of vortex core reversal (switching) triggered by vortex-hole interaction\cite{worksub}.\\

\vskip 1cm
\centerline{\large\bf Acknowledgements} \vskip .5cm
S.G. Alves is acknowledged for discussions and computational help. The authors thank CNPq and FAPEMIG (Brazilian agencies) for the financial supports.

\section{Figure Captions}

Figure 1: [Color online] Typical plots of the potential $V_{\rm eff}/A$ as a function of the vortex position, $s$. Here, we have taken $R=1000\,{\rm nm}$, $L=50 \,{\rm nm}$ (hole sizes are shown in the figures). a) centralized defects produce deeper and sharper potential wells as their sizes increase. Here, the graph corresponds to the normalized potential so that the bottons are depicted at the same normalized value. b) each eccentric hole induces a new minimum for the potential, which may be the absolute if the hole size is large enough. Since the disk center always attracts the vortex, the minimum due to an eccentric defect does take place at a given point between the disk and hole centers (here, the holes are centralized at $x=X/R=0.4$).\\

Figure 2: [Color online] The vortex equilibrium position as function of an external varying field. The linear behavior is observed without the hole. A centralized cavity deeply changes vortex dynamics as the field is continuously increased from -1 to +1, namely there appear abrupt jumps when its center is captured (released) by (from) the hole (blue dotted lines). Note also that the graph slop is greatly lowered whenever the vortex center is inside the defect (red line), exhibiting the stronger effects of the net potential (for improving visualization, we have shown only the central region of the graph, where the differences take place). Compare with related results from Ref.\cite{RahmPRL95}.\\

Figure 3: [Color online] How the gyrotropic frequency of the vortex, $\omega_G$, is influenced by the presence of a centralized hole ($\rho=100 \, {\rm nm}$). The change is slight for very large disks, but becomes pronounced as the disk radius is decreased, namely when disk and hole sizes becomes comparable.
\newpage
\section{Figures\\}

\thebibliography{99}

\bibitem{Cowburn} R.P. Cowburn, D.K. Koltsov, A.O. Adeyeye, M.E. Welland, and D.M. Tricker, Phys. Rev. Lett. {\bf 83}, 1042 (1999).

\bibitem{Raabe}J. Raabe, R. Pulwey, R. Sattler, T. Schweinb\"och, J. Zweck, and D. Weiss, J. App. Phys. {\bf 88}, 4437 (2000). 

\bibitem{Ross}C.A. Ross, M. Hwang, M. Shima, J.Y. Cheng, M. Farhoud, T.A. Savas, H.I. Smith, W. Schwarzacher, F.A. Ross, M. Redjdal, and F.B. Humphrey, Phys. Rev. {\bf B65}, 144417 (2002).

\bibitem{Shinjo} T. Shinjo, T. Okuno, R.
Hassdorf, K. Shigeto, and T. Ono, Science {\bf 289}, 930 (2000).

\bibitem{Miltat} J. Miltat and A. Thiaville, Science {\bf 298}, 555 (2002).

\bibitem{Wacho} A. Wachowiak, J. Wiebe, M. Bode, O.
Pietzsch, M. Morgenstern, and R. Wiesendanger, Science {\bf 298}, 557 (2002).

\bibitem{RahmPRL95} T. Uhlig, M. Rahm, C. Dietrich, R. H\"ollinger, M. Heumann, D. Weiss, and J. Zweck, Phys. Rev. Lett. {\bf 95}, 237205 (2005).

\bibitem{RahmAPL82}M. Rahm, M. Schneider, J. Biberger, R. Pulwey, J. Zweck, and
D. Weiss, App. Phys. Lett. {\bf 82}, 4110 (2003).

\bibitem{RahmJAP95} M. Rahm, R. H\"ollinger, V. Umansky, and D. Weiss, J. Appl.
Phys. {\bf 95},6708 (2004).

\bibitem{AfranioPRB2005} A.R. Pereira, Phys.Rev. {\bf B71}, 224404 (2005).

\bibitem{AfranioJAP2005} A.R. Pereira, J. App. Phys. {\bf 97}, 094303 (2005).

\bibitem{Capvortex}A.R. Pereira and W.A. Moura-Melo, ``{\em Vortex behavior in ferromagnetic systems with point defects: from macro to nanostructured magnets}'', in ``{\em Electromagnetic, magnetostatic, and exchange-interaction vortices in confined magnetic structures}'', edited by E. Kamenetskii, Research Signpost, Kerala, India (to appear).

\bibitem{AfranioJF} A.R. Pereira, L.A.S. M\'ol, S.A. Leonel, P.Z. Coura, and B.V. Costa, Phys. Rev. {\bf B68}, 132409 (2003).

\bibitem{FagnerPLA2004} F.M. Paula, A.R. Pereira, and L.A.S. M\'ol, Phys. Lett. {\bf A329}, 155
(2004).

\bibitem{nossoPRB2003} L.A.S. M\'ol, A.R. Pereira, and W.A. Moura-Melo, Phys. Rev. {\bf B67}, 132403 (2003).

\bibitem{RahmAPL85} M. Rahm, J. Stahl, W. Wegscheider, and D. Weiss, App. Phys. Lett. {\bf 85}, 1553 (2004).

\bibitem{RahmAPL87} M. Rahm, J. Stahl, and D. Weiss, App. Phys. Lett. {\bf 87}, 182107 (2005).

\bibitem{G-ParkCrowellPRB67} J.P. Park, P. Eames, D.M. Engebretson, J. Berezovsky, and P.A. Crowell, Phys. Rev. {\bf B67}, 020403(R) (2003). 

\bibitem{GuslienkoAPL2002}K.Y. Guslienko, V. Novosad, Y. Otani, H. Shima, and K. Fukamichi, App. Phys. Lett. {\bf 78}, 3848 (2001); Phys. Rev. {\bf B65}, 024414 (2002).

\bibitem{G-GusJAP91-02} K. Yu. Guslienko, B.A. Ivanov, V. Novosad, H. Shima, and K. Fukamichi, J. App. Phys. {\bf 91}, 8037 (2002).

\bibitem{Usov}N.A. Usov and L.G. Kurkina, J. Magn. Mag. Mat. {\bf 242-245}, 1005 (2002).

\bibitem{G-ZaspelIvanovJAP95-04} B.A. Ivanov and C.E. Zaspel, J. App. Phys. {\bf 95}, 7444 (2004).

\bibitem{G-GusPRB72-05} V. Novosad, F. Y. Fradin, P.E. Roy, K.S. Buchanan, K. Yu. Guslienko, and S.D. Bader, Phys. Rev. {\bf B72}, 024455 (2005).

\bibitem{G-GusPRL96-06}K.Yu. Guslienko, X.F. Han, D.J. Keavney, R. Divan, and S.D. Bader, Phys. Rev. Lett. {\bf 96}, 067205 (2006).

\bibitem{ComptonCrowellPRL2006} R.L Compton and P.A. Crowell, Phys. Rev. Lett. {\bf 97}, 137202 (2006).

\bibitem{HoffmanPRB2007} F. Hoffmann, G. Woltersdorf, K. Perzlmaier, A.N. Slavin, V.S. Tiberkevich, A. Bischof, D. Weiss, and C.H. Back, Phys. Rev. {\bf B76}, 014416 (2007).

\bibitem{Kuepper} K. Kuepper, L. Bischoff, Ch. Akhmadaliev, J. Fassbender, H. Stoll, K.W. Chou, A. Puzic, K. Fauth, D. Dolgos, G. Schütz, B. Van Waeyenberge, T. Tyliszczak, I. Neudecker, G. Woltersdorf, and C.H. Back, App. Phys. Lett. {\bf 90}, 062506 (2007).

\bibitem{nossoJAP2007} A.R. Pereira, A.R. Moura, W.A. Moura-Melo, D. Carneiro, S.A. Leonel, and P.Z. Coura, J. App. Phys. {\bf 101}, 034310 (2007).

\bibitem{Wysin1} G.M. Wysin, Phys. Rev. {\bf B 54}, 15156 (1996); {\em ibid} {\bf B 63}, 094402 (2001).

\bibitem{KuepperPC} K. Kuepper, private communication.

\bibitem{BahianaNanotech2007} D. Altbir, J. Escrig, P. Landeros, F. S. Amaral, and M. Bahiana, Nanotechnology {\bf 18}, 485707 (2007).

\bibitem{ShekaJMMM2007} V.P. Kravchuk, D.D. Sheka, and Y.B. Gaididei, J. Magn. Mag. Mat. {\bf 310}, 116 (2007). See also related refs. cited therein.

\bibitem{worksub} R.L. Silva, A.R. Pereira, W.A. Moura-Melo, and N.M. Oliveira-Neto, submitted to Phys. Rev. Lett. (2008).
\end{document}